\documentclass[A4paper,11pt]{article}

\usepackage[utf8]{inputenc}
\usepackage[T1]{fontenc}
\usepackage{lmodern}

\usepackage{amsfonts}
\usepackage[version=4]{mhchem}
\usepackage{amssymb}
\usepackage{amsmath}
\usepackage[usenames]{color}
\usepackage{graphicx}
\usepackage{tabularx}
\usepackage{wrapfig}
\usepackage{pdfpages}
\usepackage{subfigure}
\usepackage[small,hang,it]{caption}
\usepackage[all]{xy}
\usepackage{fancyhdr}  
\usepackage{siunitx} 
\usepackage{array}
\usepackage{graphics}
\usepackage{indentfirst}
\usepackage{multirow}
\usepackage{lscape} 
\usepackage{rotating}
\usepackage{cancel}
\usepackage[export]{adjustbox}
\usepackage[titletoc]{appendix}
\usepackage[margin=1.25in,headheight=13.6pt]{geometry}
\usepackage[percent]{overpic}

\usepackage{bibentry}
\usepackage{sectsty}
\usepackage{pdfpages}
\usepackage{authblk}
\usepackage{hyperref}

\sectionfont{\centering}

\begin{document}

\title{Evaluation of Monte-Carlo-based System Response Matrix Completeness and its Impact on Image Quality in Positron Emission Tomography}

\author{A. Hourlier}
\author{D. Giovagnoli}
\author{V. Bekaert}
\author{F. Boisson}
\author{D. Brasse}
\affil{Université de Strasbourg, CNRS, IPHC UMR 7178, F-67000 Strasbourg, France}

\maketitle


\section{Introduction}


A major strength of iterative algorithms used in positron emission tomography (PET) lies in their abilities to introduce precise models of the physics at play, which includes the statistical nature of the detection processes, and a detailed description of the radiation-matter interactions. 
The process of data acquisition by the imaging system is described in a system response model,  or system response matrix (SRM). In PET, elements of this matrix correspond to a probability of a coincident pair of $\gamma$ emitted from a certain element of the imaged volume (voxel) to be detected by the apparatus along a given pair of detection elements (e.g. a pair of scintillating crystals), or line of response (LoR).\\


The effects of the geometrical acceptance, and of the location of the different detection elements as well as the physics of interactions of the 511-keV photons in the detector need to be modelled, either analytically or through Monte-Carlo (MC) simulations. In principle, the better these effects are accounted for in the SRM, the better the quality of the final image.

However, the computation of the SRM for a given imaging device can be extremely tedious given the size of the matrix. Current and pre-clinical PET scanners reaching millimeter-level resolution over fields of view of tens of centimeters require heavy matrices for their image reconstruction. Various techniques have been proposed to compute SRMs, ranging from experimental determination of the matrix \cite{XpSRM}, analytical description of the $\gamma$ interactions \cite{AnalyticalSRM}, or full MC simulations \cite{IRISMtx}.

MC simulations are a useful tool to simulate precisely the system and the physics, in order to compute the full SRM or most of its components. 
Statistical uncertainties of the SRM elements are propagated to the final reconstructed images. For low statistic limits,  voxels contributing to a given LoR might not be represented,  leading to underestimating the geometric acceptance of some LoR.

However, noiseless SRM elements would be probably difficult to achieve and highly resource intensive. Other processes such as for example positron range, photon pair accolinearity, or absorption/diffusion along the photon path, if not accounted for, also lead to a degradation of the image quality. A balance between the SRM statistics and the available computing resources can therefore be found where the purely statistical effects are overpowered by the more systematic effects of neglecting physical processes.

The number of voxels involved for each line of response and the statistical error on each matrix element are directly dependent on the number of coincident events simulated to generate the SRM. The main goal of this paper is to first evaluate the behavior of these two parameters and secondly to estimate their impact on the overall image quality in preclinical PET imaging.

\section{Materials and methods}

For the purpose of this work, data were acquired with the IRIS PET/CT preclinical scanner, which features state of the art technology for preclinical studies on small animals and is commercialized by the Inviscan SAS company.

\subsection{Description of the IRIS tomograph}

This tomograph is composed of 16 modules arranged in two octagonal rings. Each detection module is composed of a matrix of $26 \times 27 $ (respectively along the axial and transaxial directions) scintillating Lutetium-Yttrium Oxyorthosilicate crystals doped with Cerium (LYSO$:$Ce), of dimensions $(1.6 \times 1.6 \times 12)$ mm$^{3}$ each, coupled to a $8 \times 8$ multi-anode PMT read out using an ``Anger-type logic'', allowing to locate the barycenter of the scintillation light.
The coincidence detection logic is of 1-vs-6, where a crystal can be in coincidence with any crystal in the three modules directly on the opposite side of the FoV in both rings, leading to a maximum of \num{23654592} possible LoRs. The dimensions of the FoV are 80 mm and 95 mm in the transaxial and axial directions, respectively.

\subsection{Image Reconstruction}

The coincidences collected by the IRIS are stored in histogrammed list modes files, where each LoR is associated with the number of coincidences detected by the pair of crystals, corrected for dead time and the activity reduction due to the half-life of $^{18}$F. The output files of the tomograph also include an estimation of the random count in each LoR obtained by a delayed window method \cite{RandomBelcari}

Images are reconstructed using an ordered-subset estimation maximization (OS-EM, \cite{OSEMpaper}) algorithm with 8 subsets and 8 iterations, combined with SRMs generated with various statistics (more details on the SRMs are given in subsection \ref{SRMsubsection}). Random correction is performed using the random count estimation in each LoR provided by the IRIS tomograph.

Uniformity correction is performed using a direct normalization procedure similar to the one described in \cite{IRISPEt}. A  $(90 \times 5 \times 110)$ mm$^3$ plate is filled with a $^{18}$F solution and positioned in the field of view of the tomograph. Acquisitions are performed for eight positions of the plate, rotated along the axis of the tomograph by respectively $0^\circ$, $45^\circ$, $90^\circ$, $135^\circ$, $180^\circ$, $225^\circ$, $270^\circ$,  and $315^\circ$. For each positions, coincidences involving the four modules closest to the plate are ignored. 
This protocol is then reproduced in the simulation where our SRMs are used to provide an expectation of the count rate in each LoR, assuming a perfectly uniform acquisition. The ratio of the expected count rate over the acquired rate, after random subtraction provides a uniformity correction factor that accounts for individual detection efficiency of each crystal involved in a given LoR.

We reconstruct the radiotracer distribution in a $(68\times 68\times100)$ mm$^{3}$ image volume of $(170 \times 170 \times 250)$ voxels of size $(0.4 \times 0.4 \times 0.4)$ mm$^{3}$.

A regularization is performed after each iteration using a 3D Gaussian kernel of $0.8$ mm FWHM (equivalent to a two-voxel width).

\subsection{System Matrix Generation} \label{SRMsubsection}
The SRM used to reconstruct the images were generated using the GATE simulation framework \cite{GATEref}. A uniform activity of back-to-back 511-keV photons, emitted over $4\pi$, is simulated in a cylinder of 35 mm radius and 100 mm length centered in the FoV, with no simulated attenuation.

The determination of the location of $\gamma$ interactions, and therefore of the crystals involved to define the hit LoR, is performed simulating the Anger logic of the IRIS tomograph. 

A pair of back-to-back 511-keV photons emitted from a voxel $i$ and detected by the pair of crystals defining a LoR $j$ increments the SRM element $a_{ij}$ by a value of one. This procedure produces matrices normalized to the total number of pairs of photons that interacted in the tomograph. An absolute normalization to MBq.mL$^{-1}$ of the images is performed \textit{a posteriori} using phantoms filled with known activity concentrations.

Because only a small number of the possible (voxel and LoR) couples can be explored in a reasonable computation time, storing a ``hollow'' SRM (i.e. only storing the values of non-zero matrix elements and an information on their coordinates in the matrix) allows for more manageable files, for storing and file access during image reconstruction.

Each matrix element is stored as a pair of 32-bit integer and float, respectively for the voxel index, and its contribution to a given LOR. 


In addition to these Monte-Carlo-based matrices, we also produced a matrix using a purely geometrical ray tracer, producing the list of voxels along the line between two detection crystals for each possible pair of crystals. The goal is to provide a bare-bone reference for the evaluation of the completion of the MC SRM.

We quantify the completeness of the matrix through different observables, including the averaged number of decays simulated per voxel, and how this compares to the total number of unique lines of response of the tomograph. We also report the average number of voxels contributing to a given LoR, to be compared to the same number for a purely geometrical ray tracer. The maximum number of times a given voxel contributes to a given LoR, averaged over the entire matrix, allows to estimate the granulosity of the probabilities described by the matrices. 

\subsection{Data Acquisition}

\subsubsection{Image Quality Study}

The image quality is evaluated using a pre-clinical Image Quality Phantom (IQP) following the NEMA NU 4-2008 standard \cite{NemaStd}. The IQP is arranged in three regions : 
(i) a cylinder of 30 mm diameter and 30 mm length filled with a uniform activity and in which (ii) two cold regions hollow cylinders (8 mm diameter and 15 mm length) are placed, filled with air and water. Finally (iii), a solid cold region in which five fillable rods of 20 mm length and diameters ranging from 1 to 5 mm are drilled. The phantom is filled with 19.7 mL of a $^{18}$F solution for a total activity of 3.7 MBq as recommended by the NEMA protocol. A single acquisition of 20 min is performed using the IRIS pre-clinical PET/CT scanner with the IQP in the central part of the field of view. This data will be used to evaluate the performance of the image reconstruction using SRMs of various statistics as stated above, through the measurement for each reconstruction scheme of the recovery coefficient, the spillover ratio and the signal-to-noise ratio as prescribed by the NEMA NU 4-2008 standard.

\subsubsection{\textit{In-vivo} Mice Studies}

In addition to the systematic study with the IQP, we acquired images from six mice, one injected with $^{18}$F-sodium fluoride ($^{18}$F-FNa), which highlights the osteoblastic activity, and five injected with $^{18}$F-fluorodeoxy glucose ($^{18}$F-FDG), a commonly used tracer in oncology, which allows to measure the glucose metabolism.\\

\textbf{FNa-injected Mouse}

A 27g MRL/lpr female mouse was imaged for 10 min using the IRIS preclinical PET/CT scanner. The mouse had previously been injected with 19.1 MBq of $^{18}$F-FNa 112 min prior to the beginning of the imaging. The estimated activity at the start of the acquisition was 9.7 MBq. Anesthesia was induced during the biodistribution with a $2\%$ isofluorane mix in medical air. The mouse was sacrificed prior the imaging PET/CT procedure. CT acquisition (80 kV, 0.9 mA) was performed for anatomical correlation.
From the reconstructed CT volume, ten vertebra were manually segmented and used to extract the mean value inside each vertebra of the different reconstructed PET volumes. An additional volume of interest (VOI) was manually derived to take into account the values inside the central spine region along the first six vertebrae.\\

\textbf{FDG-injected Mice}

PET imaging was performed 7 days after 4T1 breast cancer cells allograft on five female BALB/c mice. 4.8 ± 0.5 MBq of 18F-FDG were injected into mice tail vein. All PET/CT acquisitions were performed on an IRIS PET/CT device (Inviscan, France). For each mouse, a 10 minutes static acquisition was performed at 60 minutes after radiotracer injection followed by a CT acquisition.  At the end of the PET/CT exam, each mouse was sacrificed. After excision of the tumor, the specimen was weighed and placed in an automatic gamma counter (Hidex, Turku, Finland) for a 1 minute acquisition. The count obtained was converted to absolute activity using a calibration curve and then corrected for radioactive decay to an activity value arbitrary taken at the time of animal sacrifice. The results are expressed as a percentage of injected dose per gram of tissue (\%ID/g). For a semiquantitative assessment of tumoral uptake on the reconstructed PET images, VOIs were drawn according to the shape of each tumor to assess their activity and to estimate the impact of each SRM on the measured percentage of injected dose per gram of tumor (\%ID/g).

All animal experiments were performed in accordance with the European Institutes of Health Guidelines regarding the care and use of animals for experimental procedures. The study protocol was approved by the Alsace Regional Ethics Committee for Animal Experimentation (Approval ID: APAFIS ${\#}$1531 and ${\#}$15255 for FNa and FDG experiments)

\section{Results and Discussion}

\subsection{System Matrix Generation}

The total activity generated per voxel of the simulated uniform cylinder for each of the seven matrices we evaluated is summarized in Table \ref{tab:statTab}.

\begin{table}[!h]
    \centering
    \caption{Summary of the statistics used to generate the system response matrices  with which the various datasets have been reconstructed.}
    \small
    \begin{tabular}{m{4.5cm} | c c c c c c c c}
         & (a) & (b) & (c) & (d) & (e) & (f) & (g) & ray tracer \\
        \hline
         Simulated $\gamma$-pairs per voxels  ($\times 10^{3}$)       &  5.9   &  11.8  &  \num{23.6}  &  \num{47.1}  & \num{94.0}  & \num{188.2}  & \num{376.4}  & --- \\
         Simulated $\gamma$-pairs per voxels per LoR ($\times 10^{-3}$) & 0.01 & 0.2 & $0.39$ & $0.8$ & $1.6$ & $3.2$ & $63$ & --- \\
         Average nb. of voxels contributing to a same LoR & 42 & 75 & 131 & 231 & 410 & 73 & 1296 & 193 \\
         Maximum nb. of times a single voxel contributes to the same LoR        &    4     &   5      &   6    &    8   &       10 &    13    &     20   & 1\\
         File size (GB)                     & 10.0   &  20.0  & 39.6 & 78.0 & 151.0 & 286.1 & 519.4 & 27.1
         
    \end{tabular}
    \label{tab:statTab}
\end{table}

\begin{figure}[!h]
    \centering
    \includegraphics[width = 0.75\textwidth]{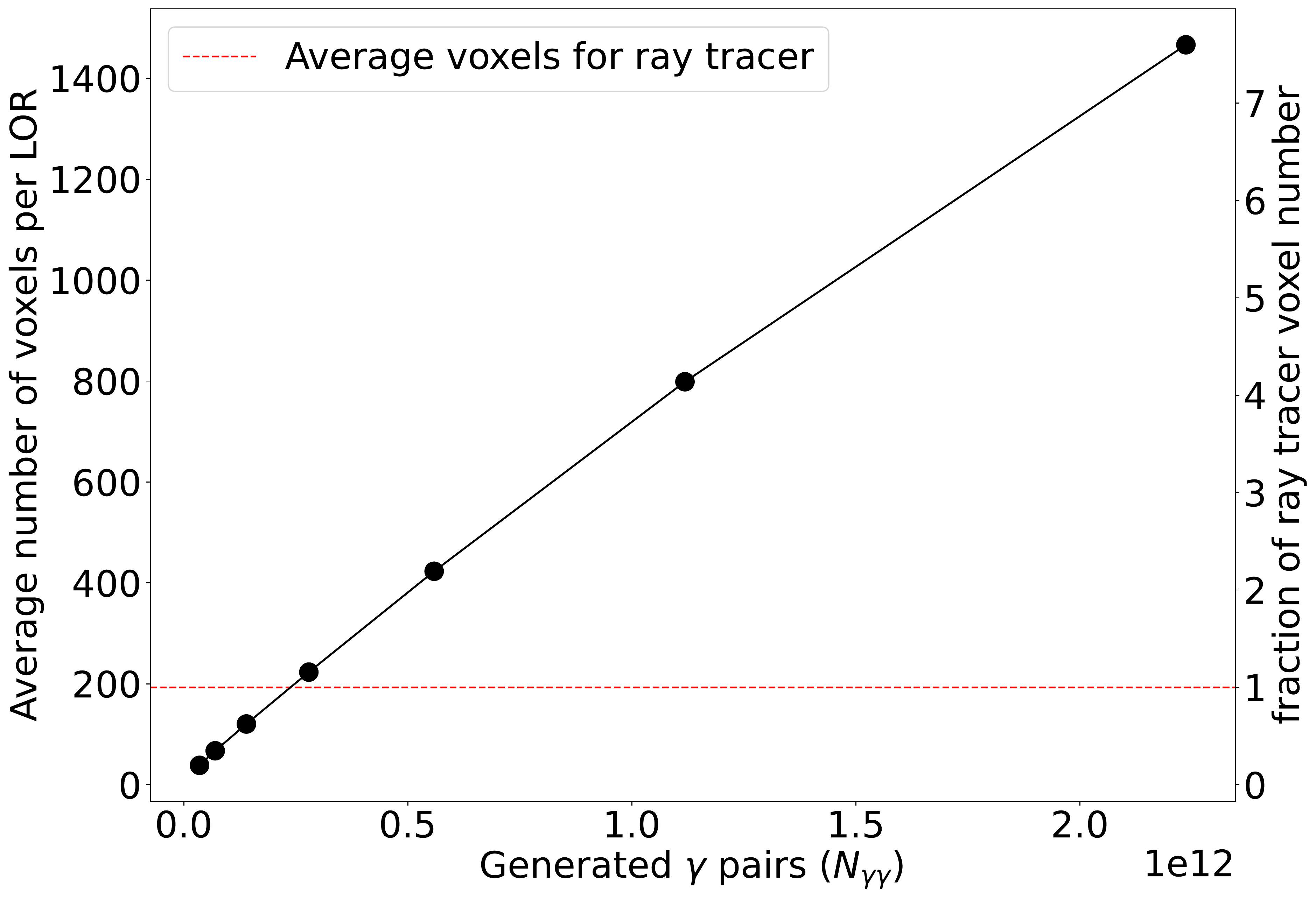}
    \caption{The average number of voxels per line of response as a function of $\gamma$ pairs generated in the field of view compared to the average number of voxels per line of response in the case of a ray-tracer operator approach.}
    \label{fig:NbVoxLOR}
\end{figure}

The average number of voxels per line of response as a function of $\gamma$ pairs generated in the field of view of the PET imaging system is presented in Figure \ref{fig:NbVoxLOR}. If we project all the voxels associated to a given LOR on a cross sectional plane of this LOR, while a ray tracer based approach gives one centered pixel, the SRM approach produces a probability density function with an increasing number of involved pixels as the statistics of the matrix increases. 
The additional voxels around that line are introduced by the diffusion in the detector. The matrices (a) to (c) are still less complete than the ray tracer and noticeable gaps in the voxel distribution along the LoR are to be expected.  Increasing the completeness of the matrix would therefore  equivalent to increasing of the sampling of the diffusion kernel.

\begin{figure}[!h]
    \centering
    \includegraphics[width = 0.75\textwidth]{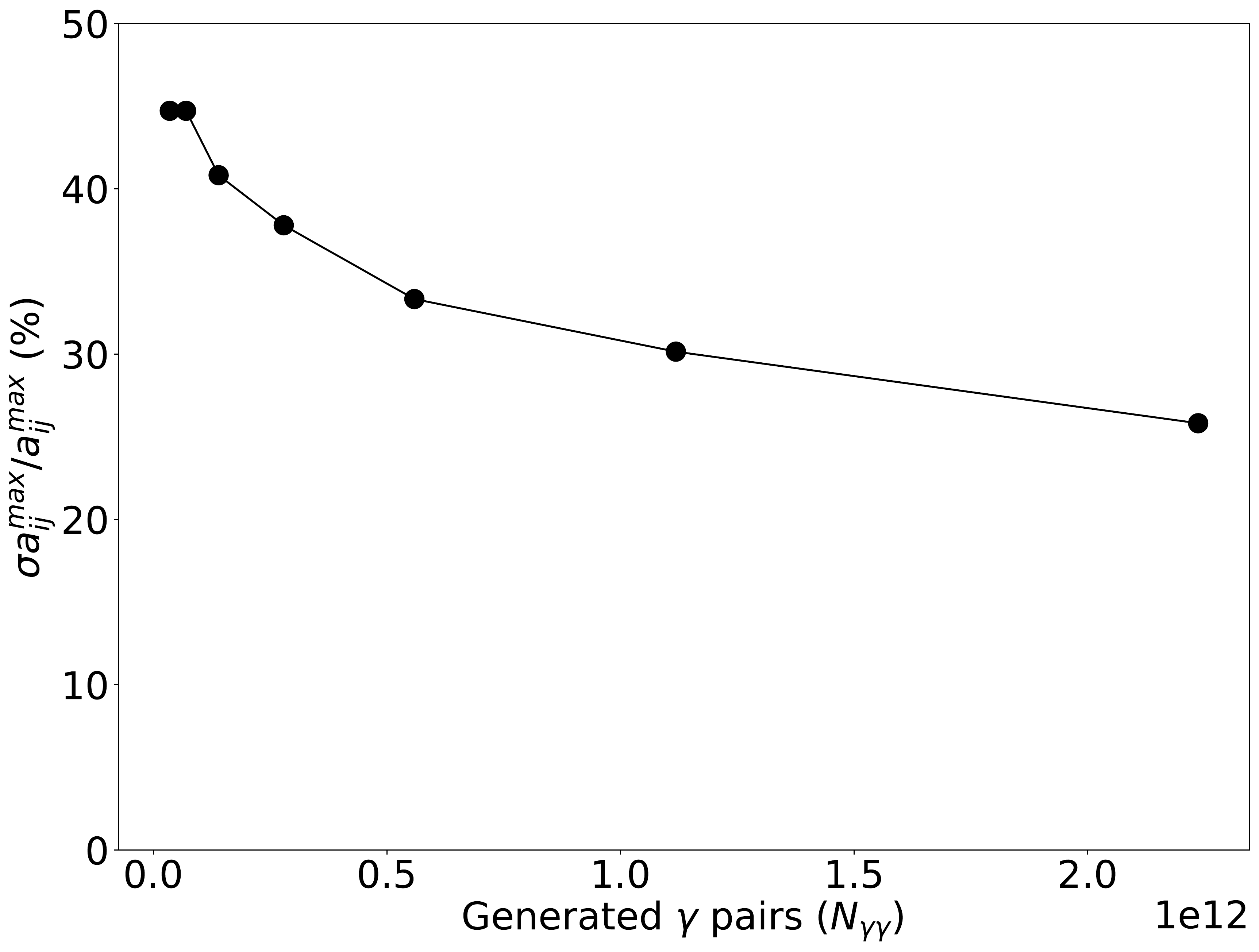}
    \caption{The relative uncertainty on the values of the SRM elements with the highest probabilities for each lines of response, averaged over all LOR, as a function of the SRM statistics.}
    \label{fig:SRMelementsError}
\end{figure}

Each element of the SRM is the probability that a coincident event emitted from a particular voxel is detected along a given LOR. This probability value depends mainly on the solid angle defined by the LOR, and the intrinsic detection efficiency of the detector. Figure \ref{fig:SRMelementsError} represents the relative uncertainty of the SRM elements with the highest probability value, i.e. along the LOR central line. The behaviour appears mostly statistically driven,  our heaviest matrix of $\sim 520$ Go reaches a better than $30 \%$ uncertainty.

\subsection{Image Quality Study}

\begin{figure}[!h]
    \centering
    \subfigure[]{\includegraphics[width = 0.44\textwidth]{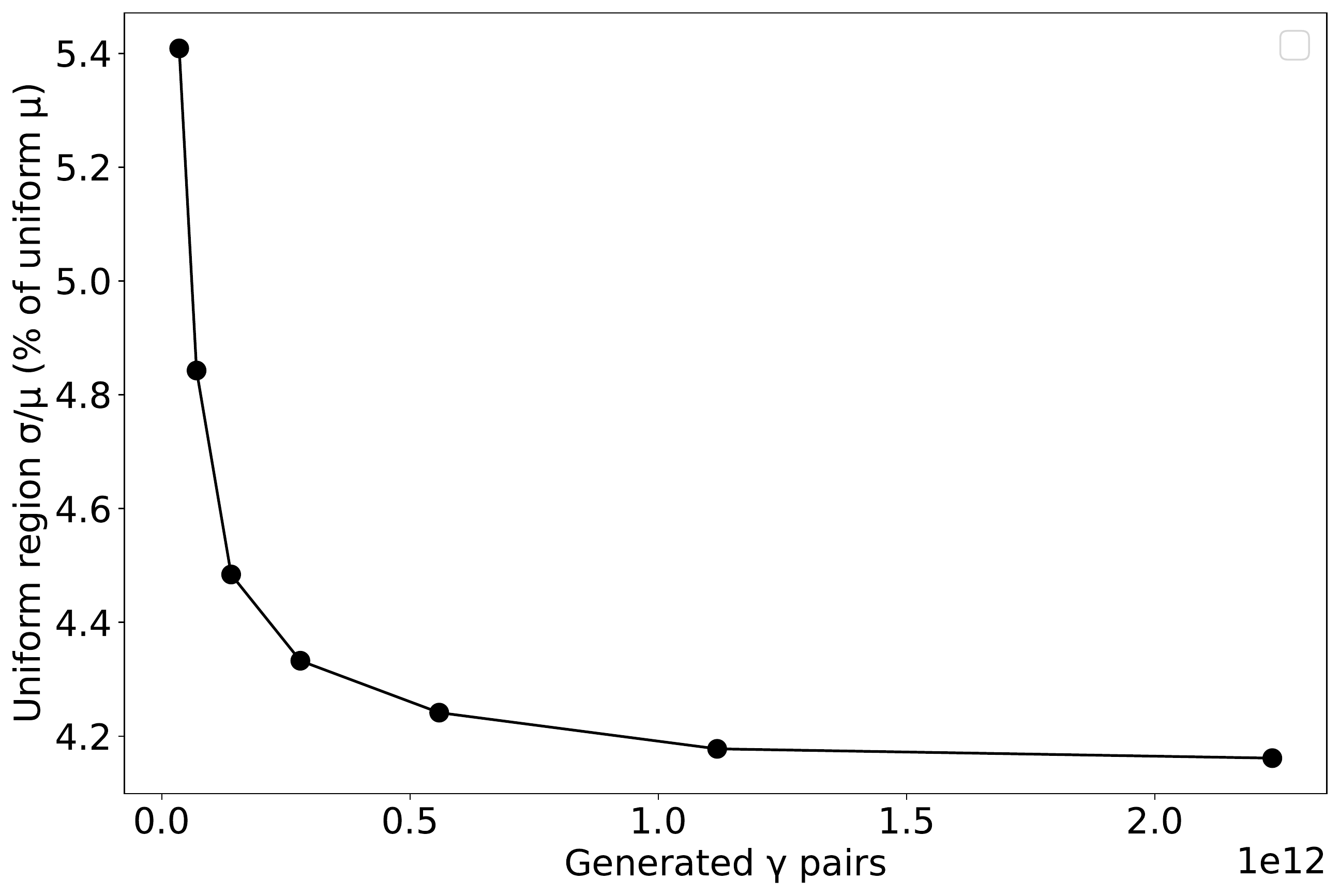}}
    \subfigure[]{\includegraphics[width = 0.55\textwidth]{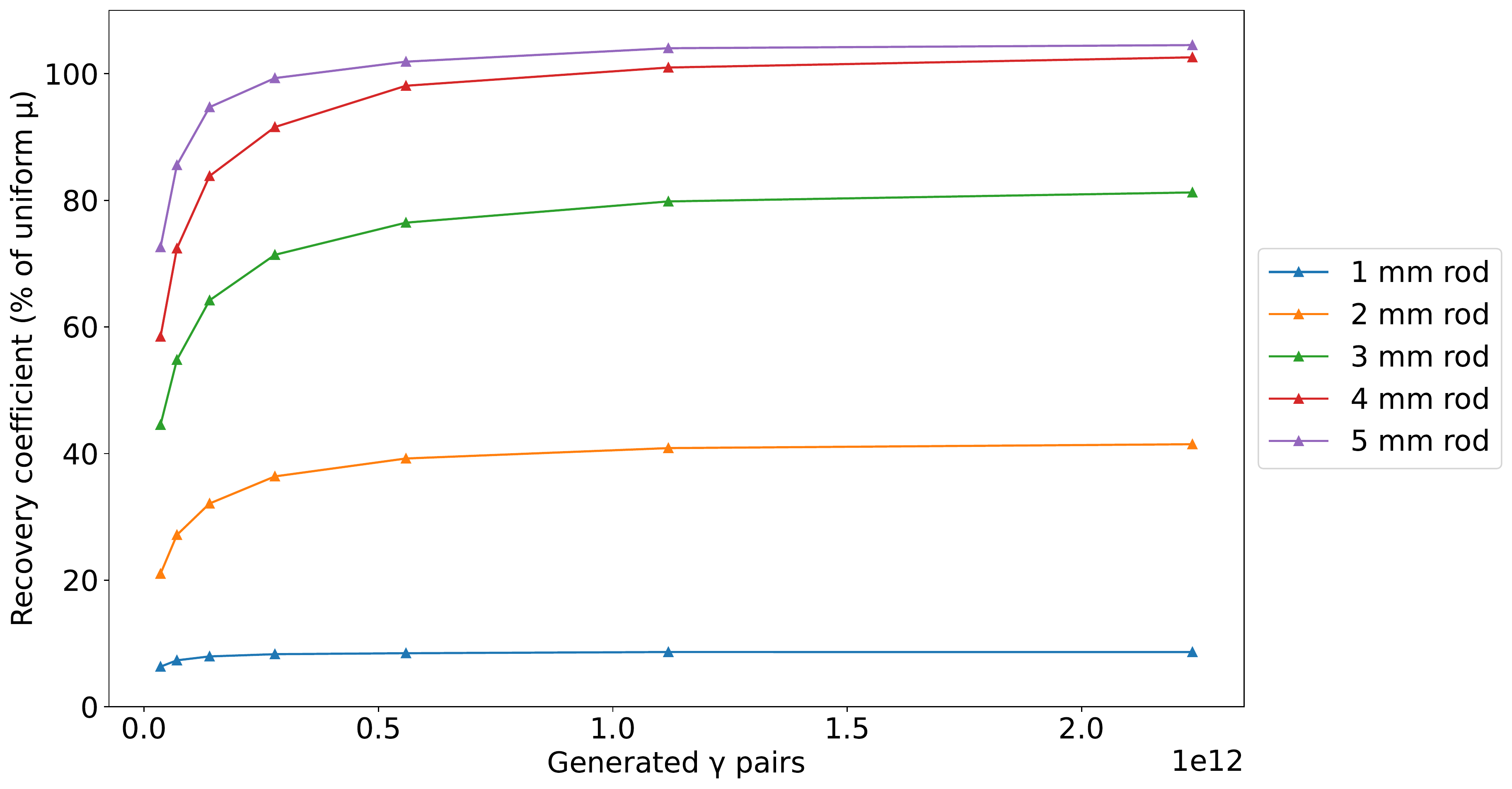}}
    
    \caption{Uniformity (a) and recovery coefficients (b) measurements following the NEMA NU-4 2008 standard as a function of SRM statistics.}
    \label{fig:recovCoeff}
\end{figure}

The reduction of the SRM uncertainty directly translates into an improvement of the uniformity of the central region of the IQP phantom (Figure \ref{recovCoeff},a). Similarly, the recovery coefficients improve with the SRM statistics (Figure \ref{fig:recovCoeff},b), but reach a plateau value faster than the uniformity, suggesting a non-statistical limitation becoming dominant. One explanation can be the regularization process during the iterative reconstruction procedure. The same behaviour can be observed for the spill over ratio measurement (data not shown).

\subsection{\textit{In-vivo} Mice Studies}

\begin{figure}[!h]
    \centering
    \subfigure[]{\includegraphics[width =  0.3\textwidth]{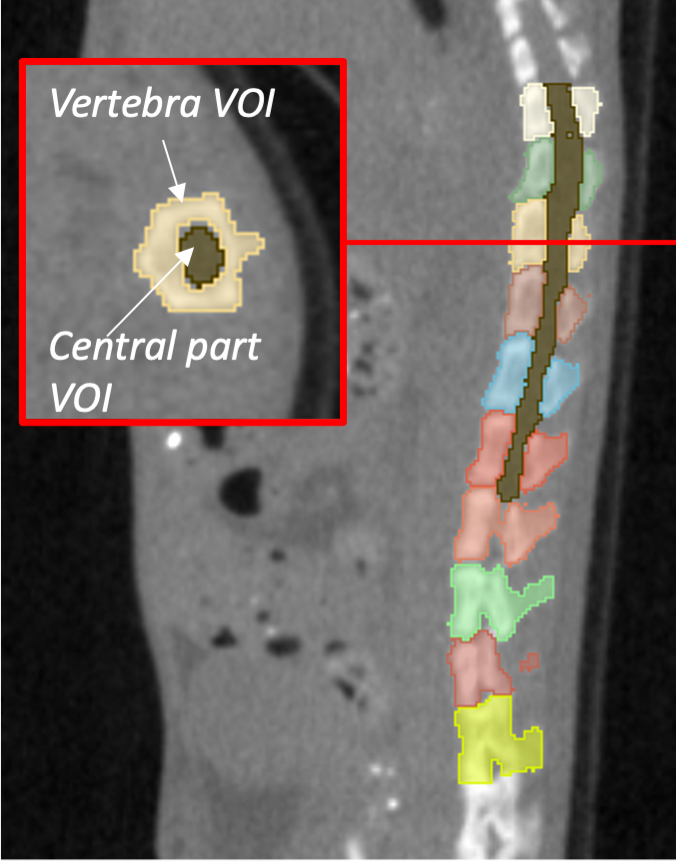}}
    \subfigure[]{\includegraphics[width = 0.65\textwidth]{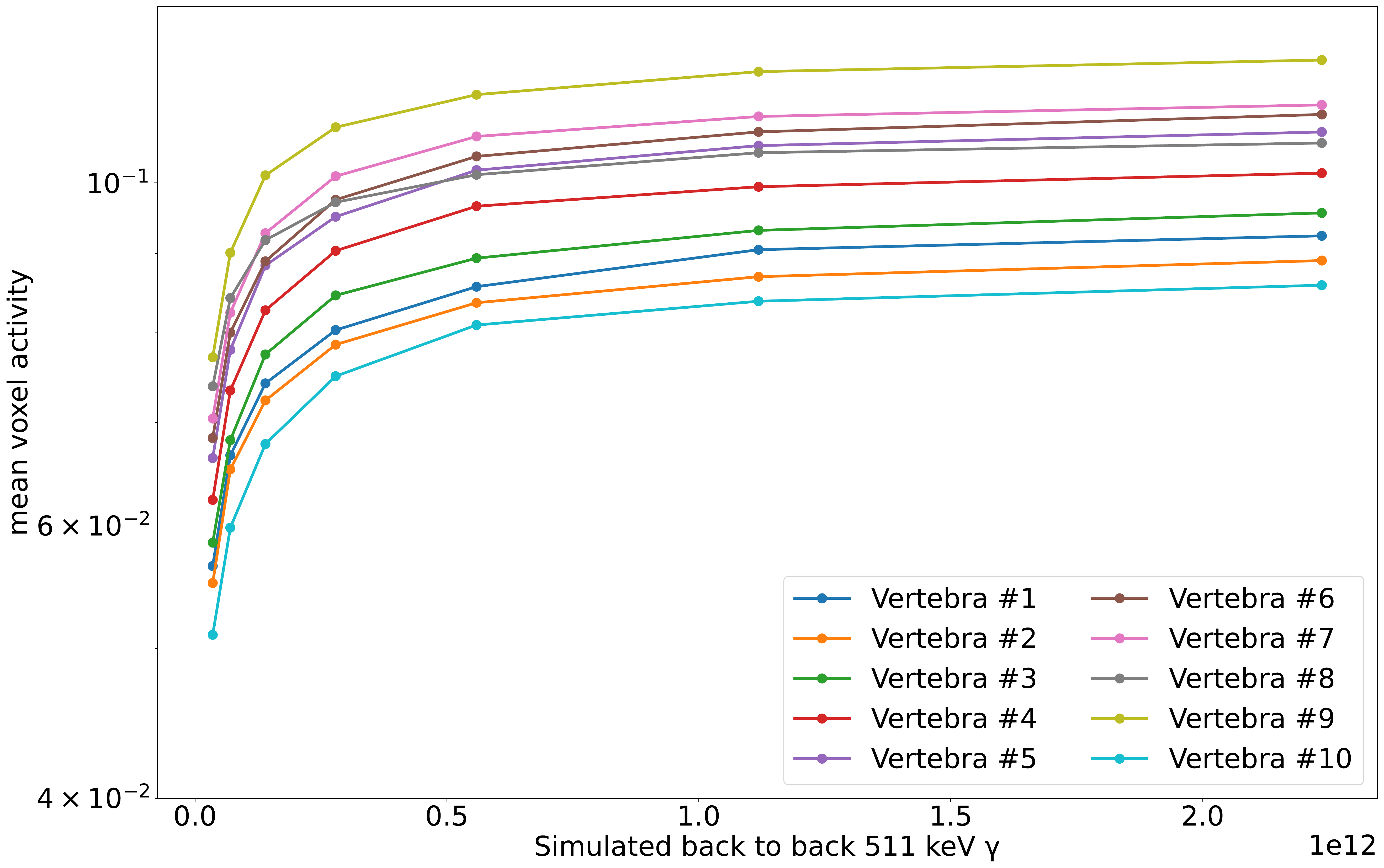}}\\
    \subfigure[]{\includegraphics[width = 0.8\textwidth]{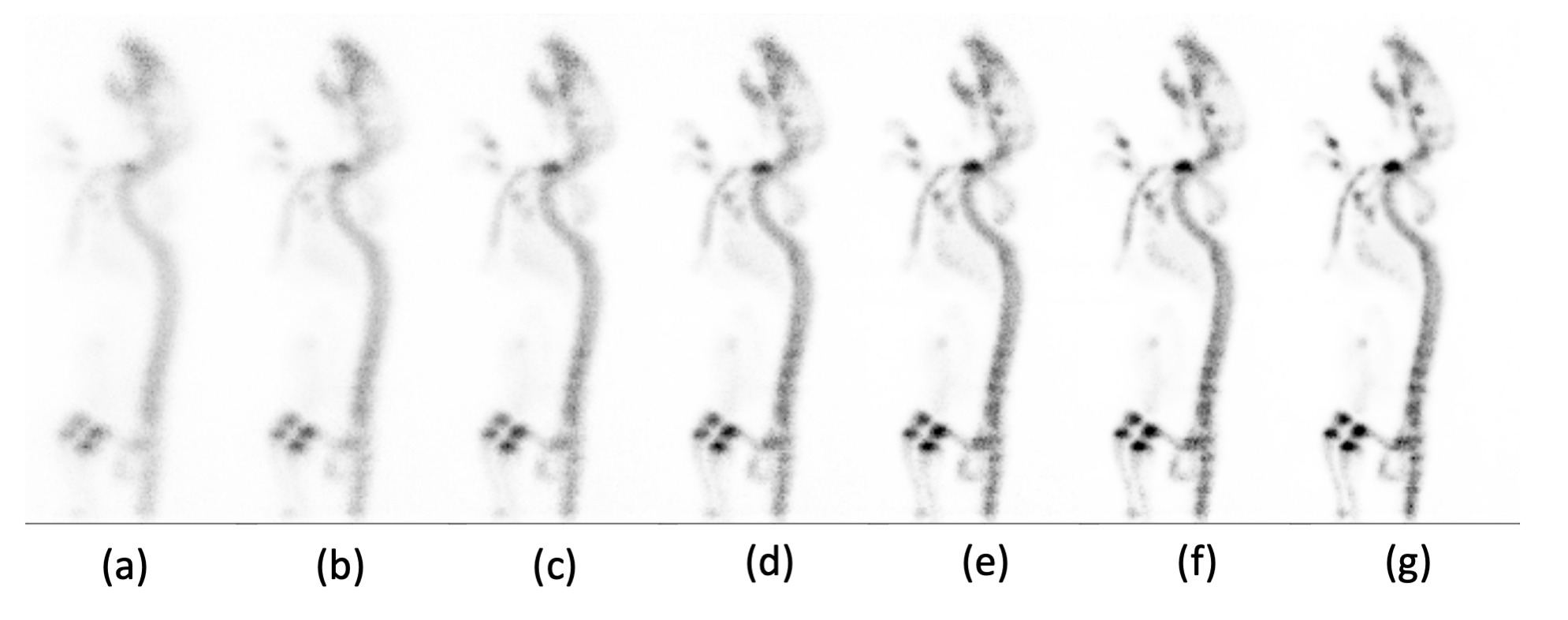}}
    \caption{(a) CT sagittal view of a FNa injected mouse used to segment the various volumes of interest. Each colored region corresponds to a hand-segmented mask based on the reconstructed CT volume. 10 vertebrae are considered, numbered 1-10 from tom to bottom, and an additional volume is segmented corresponding to the central part of the spine for the first six vertebrae\\
    (b) Average voxel activity in each individual vertebra as a function of SRM statistics.\\
    (c) Evolution of the Maximum Intensity Projection of the FNa biodistribution with SRM statistics.}
    \label{fig:FNaCTsegmentation}
\end{figure}

The FNa uptake on each vertebra on a single dataset is obtained by applying the masks  highlighted in Figure \ref{fig:FNaCTsegmentation}(a). The behaviours of the recovery coefficients with the SRM statistics are observed as well on the extracted activity uptakes in each vertebra, as shown in Figure \ref{fig:FNaCTsegmentation}(b). The apparent spatial resolution and the detectability improve as well as the the activity quantification (Figure \ref{fig:FNaCTsegmentation}(c)). 

\begin{figure}[!h]
    \centering
    \subfigure[]{\includegraphics[width = 0.95\textwidth]{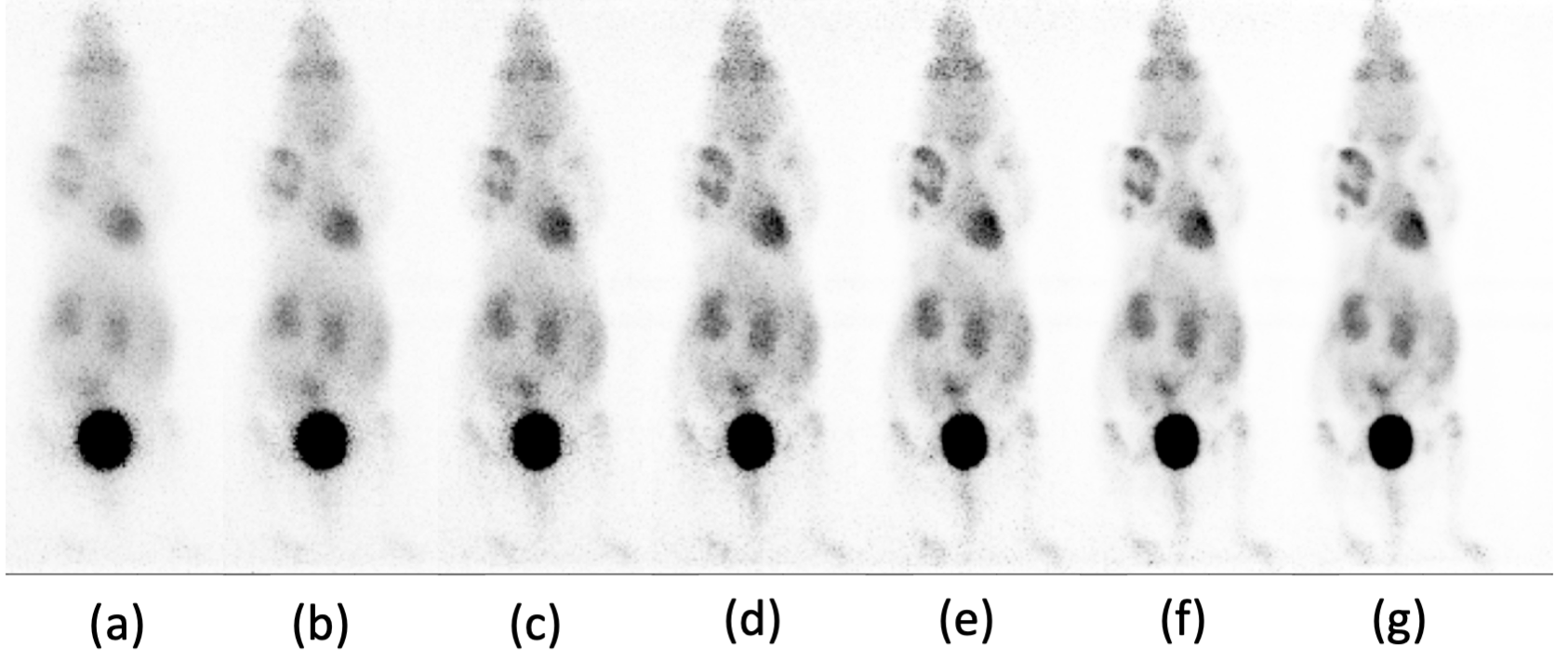}}\\
    \subfigure[]{\includegraphics[width = 0.8\textwidth]{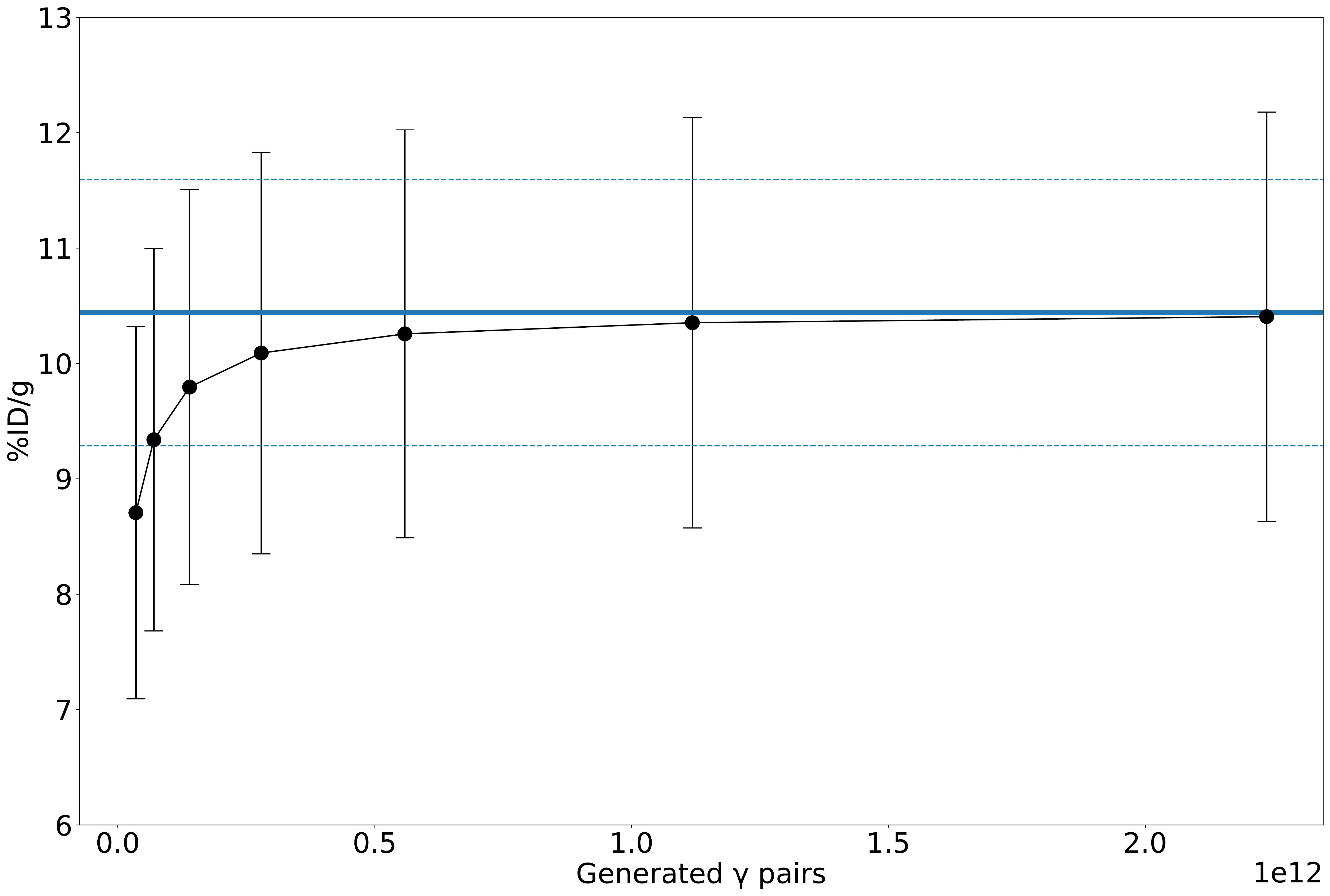}}
    
    \caption{(a) Evolution of the Maximum Intensity Projection of the FDG biodistribution with SRM statistics.\\
    (b) Improvement of the the average injected dose percentage per gram ($\%$ID/g) of breast tumors of 5 mice as a function of SRM statistics. The solid and dashed blue lines represent the dispersion of ex-vivo activity measurements of the tumors.}
    \label{fig:FDGmice}
\end{figure}

The maximum intensity projection of the reconstructed FDG biodistribution (Figure \ref{fig:FDGmice},a) as a function of SRM statistics, illustrates the impact of the SSRM completeness on the overall image quality. This qualitative improvement is also quantitatively observed when the percentage of injected dose is derived. The error bar in Figure \ref{fig:FDGmice}(b) is only due to the variability of the FDG uptake on each tumor, also observed in the \textit{ex-vivo} measurements.

\pagebreak

\section{Conclusion}


Our results show the direct impact of the statistical variance of the Monte-Carlo generated System Response Matrices used in iterative reconstruction algorithms on image quality. 

The incremental improvements of quantification and image quality are to be balanced against the computing requirements to generate and more importantly to handle matrix files during the image reconstruction procedure.

However, the statistics needs to be high enough to better describe the detector physics than a purely geometrical ray-tracer.

\bibliographystyle{unsrt}
\bibliography{statMtxBib.bib}

\end{document}